\documentclass[aps,pra,twocolumn,superscriptaddress]{revtex4}
\usepackage{mathtools}
\usepackage{graphicx}
\usepackage{epstopdf}
\usepackage{lmodern}
\usepackage{xfrac}
\usepackage{natbib}
\usepackage[breaklinks]{hyperref}
\usepackage{verbatim}
\usepackage{mathtools}

\begin{document}
\title{Countering detector manipulation attacks in
  quantum communication through detector self-testing }

\author{Lijiong~Shen}
\affiliation{Centre for Quantum Technologies, National University of Singapore, 3 Science Drive 2, Singapore 117543}

\author{Christian~Kurtsiefer}
\affiliation{Centre for Quantum Technologies, National University of Singapore, 3 Science Drive 2, Singapore 117543}
\affiliation{Department of Physics, National University of Singapore, 2 Science Drive 3, Singapore 117551}

\email[]{christian.kurtsiefer@gmail.com}
\date{\today}
\begin{abstract}
In practical quantum key distribution systems, imperfect physical devices open
security loopholes that challenge the core promise of this technology. Apart
from various side channels, a vulnerability of single-photon detectors to
blinding attacks has been one of the biggest concerns, and has been addressed
both by technical means as well as advanced protocols. In this work, we
present a countermeasure against such attacks based on self-testing of
detectors to confirm their intended operation without relying on specific
aspects of their inner working, and to reveal any manipulation attempts.  We
experimentally demonstrate this countermeasure with a typical InGaAs
avalanche photodetector, 
but the scheme can be easily implemented with any single photon detector.
\end{abstract}

\maketitle

\textit{Introduction --} 
Quantum key distribution (QKD) is a communication method that uses quantum
states of light as a trusted courier such that any eavesdropping attempt in this
information transmission is revealed as part of the underlying quantum physics
of the measurement process on the states~\cite{BB84,ekert:91,Gisin}. While the
basic protocols are secure within their set of assumptions, practical QKD
systems can exhibit vulnerabilities through imperfect implementation
of the original protocol scenarios, through imperfect preparation and detection
devices, or through side channels that leak information out of the supposedly
safe perimeter of the two communication
partners~\cite{Scarani2009,Scarani2014,Xu2020}. Families of such vulnerabilities have been
identified and addressed through technical measures and advanced
protocols. Examples are the photon number splitting attacks where single
photons were approximated by faint coherent
pulses~\cite{Brassard2000,Lutkenhaus2000}, Trojan horse
attacks~\cite{Vakhitov2001,Gisin}, various timing
attacks~\cite{Qi2007,Lamas-Linares2007,Zhao2008} and classes of information
leakage into parasitic degrees of freedom.

\begin{figure}
  \includegraphics[width=\linewidth]{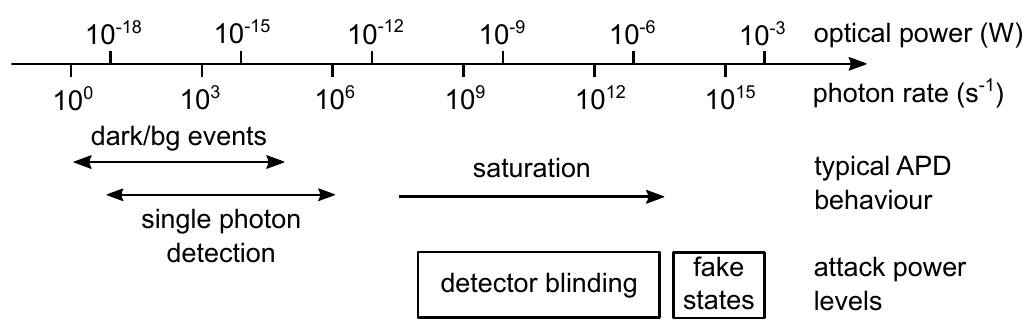}
  \caption{\label{fig:powers}Single photon avalanche photodidode properties
    underlying a blinding/fake state attack. At light levels below
    $10^{-12}$\,W, these 
    devices respond with detection events that can be used to identify single
    photons. At higher power levels, they saturate and can eventually brought
    into a blinded mode where they are not susceptible anymore to additional
    single photons. Very bright short pulses of light (``fake states'') can
    lead to a detector response that is indistinguishable  from the single
  photon response at low light levels. Photon rate/power level
  scaling is shown for a wavelength of 1300\,nm.} 
\end{figure}

Perhaps the most critical vulnerability of QKD sys-
tems is the detector blinding / fake state attack fam-
ily on single-photon detectors~\cite{Makarov2009}. This attack has been experimentally demonstrated
to work for detectors based on avalanche photodiodes and superconducting
nanowires~\cite{Lydersen2010,Lydersen2011,Goltsman2019}, and allowed to
completely recover a key generated by QKD without being noticed by the error
detection step in a QKD implementation~\cite{Gerhardt2011}.
The attack is based on the fact that these single photon detectors can be
blinded by a macroscopic light level into not giving any response, while an
even stronger light pulse or a recovery event from a blinded state could
create an output signal from the blinded detector that emulates a
photon detection event~\cite{Makarov2009} (see Fig.~\ref{fig:powers}).
This vulnerability can be
exploited by carrying out an undetected man-in-the-middle attack, where
an eavesdropper intercepts photon states carrying the information, measures
the quantum state in a basis of his/her choice,  and copies the measurement
results into the photon detector of the legitimate receiver with macroscopic
powers of light. 

Various countermeasures against the detector control attack have been
suggested and implemented. One class of countermeasures addresses technical
aspects of the detectors. Examples are using more than one
detector or a multi-pixel detector for one measurement
basis~\cite{Honjo2013,FerreiraDaSilva2015,Wang2016,Gras2021}, including a
watchdog detector for the blinding light~\cite{Lydersen2010,Dixon2017}, 
effectively varying the detector
efficiency at random timings~\cite{Lim2015,Qian2019}, and
carefully monitoring the  photocurrent or breakdown status of the
detector~\cite{Yuan2010,Yuan2011} to identify a detector manipulation.
However, most of these countermeasures have operational drawbacks. For
example, additional single photon detectors significantly increase the overall
cost and complexity, and beam splitters in the receiver for watchdog detectors
introduce additional optical losses. 
Varying the efficiency frequently to get enough statistics to identify 
the blinding attack could significantly affect
the QKD bit rate, and changing the detector operation condition or
monitoring its state increases the complexity of the electronic circuitry
around the single photon detectors. Such countermeasures may also introduce
additional vulnerabilities that may be exploited in an arms race style~\cite{Cnt-cnt-Makarov}.

An elegant countermeasure on the protocol level is provided by the so-called
measurement-device independent quantum key distribution
(MDI-QKD)~\cite{MDIQKD:2012}, which further developed the idea of
device-independent QKD where a photon pair source
can be made public or even controlled by an eavesdropper~\cite{DIQKD:2009} to
a scenario where the detectors receiving single photons (or approximations thereof) can be public, or controlled by an eavesdropper.
The scheme has been
demonstrated experimentally several times by
now~\cite{MDIexp1,MDI200km,MDI400km,MDIasym}. It requires a pair of single
photons (or weak coherent pulses) from two communication partners without a
phase correlation to arrive within a coherence time 
on a Bell state analyzer, where single
photon detection is carried out, and the result is published. This requires a
matching of emission times and wavelengths of two spatially separated
light sources with both communication partners.

The MDI-QKD approach counteracts any active or passive attack on single
photon detectors, as their result need not to be private anymore. The
communication partners can simply test if the detectors were performing single
photon detection through a error detection process similar to the original QKD
protocols.

In this work, we present a method of testing the proper operation of single
photon detectors in a QKD scenario that does not require the synchronization
of light sources like in the MDI-QKD approach, while also not touching the
specific detector mechanism. It brings the idea of self-testing of quantum
systems~\cite{Mayers1998QuantumCW,vanDam:2000,Supic2020selftestingof} to
single photon detectors that can remain black boxes. We use a 
light emitter (LE) under control of a legitimate communication partner that
is weakly coupled to its single photon detector
for this self-testing.
When the single photon detector is under a blinding attack, it is insensitive to
low-intensity light fields 
used for quantum
key distribution. Thus, by turning on the LE at times not predictable by an
eavesdropper, ``salt'' optical detection events are generated in the
detector when it operates normal, while it does not react to the test light when
blinded. Complementary, the test light intensity can be raised to blinding
levels of the photodetector, which thereby is desensitized to legitimate
single photons. Registration of any detector events under self-blinding then
suggests the presence of fake state events.

\textit{Self-testing strategy --}
In a generic QKD system, a transmitter generates photons containing
quantum information in either polarization or time encoding, and sends them
through an optical path (``quantum channel'') to a receiver. Therein, a
measurement basis choice is made either through passive or active optical
components, and the light arriving from the quantum channel is directed to
single photon detectors.
In a blinding/fake state attack, an eavesdropper measures a photon in the
quantum channel, and copies the result into the corresponding photon detector
of the legitimate receiver using blinding and fake state light levels. For
detector testing, a light emitter (LE) in the receiver is controlled by a
random number generator and weakly coupled to the single photon detectors.

\begin{figure}
  \includegraphics[width=0.9\linewidth]{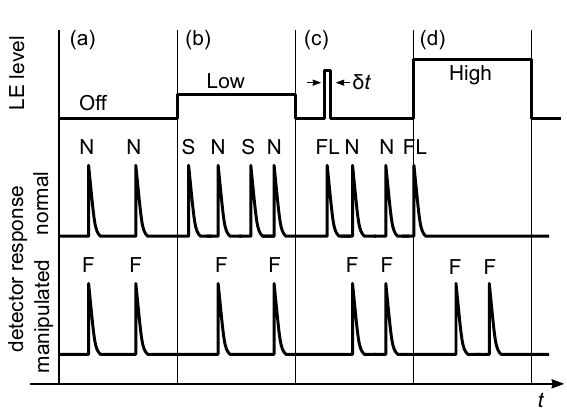}
  \caption{\label{fig:st-protocol}
    Detector self testing. Top trace: light level
    of the light emitter LE, middle trace: normal detector response (no
    manipulation), lower trace: detector response under manipulation.
    Detector events are classified as normal (N),  salt (S), ``fake'' (F), and
    flag (FL) signals. Segment (a) shows responses without
    self-testing, (b) with low LE power generating salt events, (c)
    with occasional test pulses at medium power, (d) with high LE power to
    self-blind the detector.
    }
\end{figure}

An unblinded single-photon detector generates events due to photons from the
legitimate source or the background (labeled ``N'' in
Fig.~\ref{fig:st-protocol}(a)).  The brightness of the legitimate source, the
transmission of the quantum channel, the efficiency of the single photon
detectors, and the detector dark count rate determine the average number
$\bar{n}$ of the photon-detection events registered in a time interval $T$. An
eavesdropper would choose a rate of ``fake'' detection events (labeled ``F''
in Fig.~\ref{fig:st-protocol}(a)) similar to normal QKD
operation to prevent detecting the attack by monitoring photon detection
statistics. 

We illustrate three different examples of detector self-testing to detect
detector manipulation attacks.

In the first one, the legitimate receiver switches occasionally the light
emitter LE to a low light level for a test time interval $T$ at a random timing
unpredictable by an eavesdropper, while it is off for the rest of the time. 
In the test interval, an unblinded detector would see an increase of detector
events above $\bar{n}$ due to additional salt events (``S'' in
Fig.~\ref{fig:st-protocol}(b)).  The legitimate receiver has complete control
of the light emitter to make excess photon detection events statistically
detectable in the probe interval $T$. A single photon detector under blinding
attack would be insensitive to the low light levels of LE, so only
detector events generated by positive detector manipulations like fake states
would be registered (labeled ``F'' in Fig.~\ref{fig:st-protocol}(b)). A
statistically significant presence of salt events in a time interval $T$ would
therefore allow to sense a negative detector manipulation e.g. through
blinding. Note that the test interval $T$ does not need to be
distributed contiguously in time.

This leads to a second self-testing example, which turns on the light emitter
for a short pulse time interval $\delta t$  at a random timing and with a high
enough energy (a few photons) to cause a detection event with almost
unit probability in an unblinded single-photon detector.
A blinded detector is again insensitive to such a short optical pulse as
long as the light level is way below the fake state threshold. In this
situation, detecting a single flag event can witness a non-blinded detector
(see Fig.~\ref{fig:st-protocol}(c)).

The third self-testing example uses the light emitter in the receiver to
locally blind the detector. The typical power necessary to blind an APD is on
the order of a few nW, which can easily be accomplished by weakly coupling
even faint light sources like LEDs.  Detection events caused by single photons from the legitimate source
will be suppressed by the local blinding light. In absence of a negative
detector manipulation (e.g. detector blinding),  the intense light at the
onset of the self-blinding period will almost deterministically create a flag
event in the detector, which then remains silent during the rest of
the self-blinding interval (see Fig.~\ref{fig:st-protocol}(d)).
However, any positive detector manipulation 
will overrule the local blinding, and cause a false detection event.
Both the initial flag event and any possible later event can be easily
checked.
This method only requires a small number of
registered events in a time interval $T$ to discover both a
negative and positive detector manipulation attack.

A detector event could also be triggered when the detector
recovers from a (remote) blinding exposure~\cite{Tanner2014}. Local blinding
will suppress such ``fake'' detector events, so they may not get noticed by
looking for signals under local blinding. However, in such a case, the
flag event will also be suppressed. Therefore, a combination of checking
for detection events during self-blinding and looking for a flag event is necessary to identify such an attack.

\begin{figure}
  \includegraphics[width=0.94\linewidth]{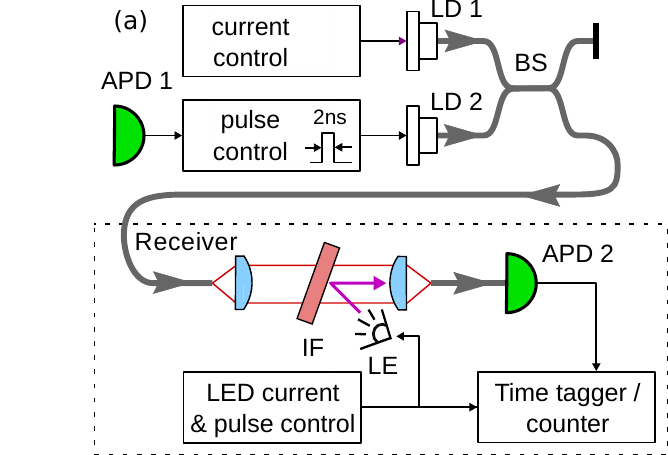}
  \includegraphics[width=1\linewidth]{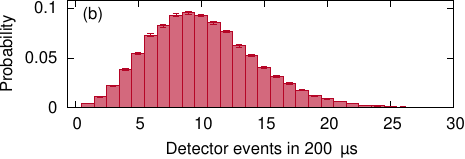}
   \caption{\label{fig:demosetup}
    (a) Setup to demonstrate detector self-testing. Light from a cw
    laser diode (LD1) and pulsed laser diode (LD2), both around 1310\,nm, is
    combined in a fiber beamsplitter (BS) to simulate different
    illumination scenarios. Besides the single photon InGaAs detector
    APD2, the receiver contains an LED (940\,nm) as a light emitter (LE)
    for local testing of APD2. An interference (IF) filter prevents leakage of
    LE light out of the receiver. (b) Distribution of photodetection events in
  a time window of $T=200\,\mu$s under ``normal'' operation under
  illumination of the detector with a low power level from LD1.}
 \end{figure}
 
\textit{Experimental results --}
We demonstrate our countermeasure with a single-photon detector
commonly used in quantum key distribution
which is susceptible to manipulation
attacks (see Fig.~\ref{fig:demosetup}(a)). Light that simulates legitimate
quantum signals and provides the larger power levels required for
detector manipulation is generated by combining the output of a continuous wave
(cw) laser diode (LD1) with light from a pulsed laser diode (LD2) on a fiber
beam splitter (BS). The 2\,ns long bright fake states from LD2 can be
emitted upon detection events from an auxiliary avalanche photodetector
(APD1) to emulate a credible (Poissonian) event distribution.
On the receiver side, the light from the quantum channel passes through an
interference filter (IF) before it is focused onto the main photodetector
(APD2), a passively quenched InGaAs device (S-Fifteen Instruments IRSPD1) with a
maximal count rate of $5\times10^5\,\text{s}^{-1}$ and a dark count rate of
$7\times10^3\,\text{s}^{-1}$. The light emitter (LE) for detector self-testing
is a light emitting diode with a center wavelength of 940\,nm (Vishay
VSLY5940), which is reflected off the IF (acting as a
dichroic beam splitter) onto APD2.

For the demonstration, we consider an event rate of
$\approx5\times10^4\,\text{s}^{-1}$ at APD2, which is about an order of magnitude
below the maximal detection rate to not reduce the detector efficiency
significantly. Figure~\ref{fig:demosetup}(b) shows a histogram of detection
events in a time interval of $T=200\,\mu\text{s}$ generated by choosing
an appropriate light level of LD1. The result with a mean photodetection number
$\bar{n}\approx10$ differs slightly from a  Poisson distribution since the
detector has an after-pulse possibility of about 40\,\%.
To implement a detector manipulation with the same event characteristic, we
elevate the optical output power of LD1 to 500\,pW, the minimal power to
completely blind detector APD2. Fake states that emulate photodetection events
in APD2 are generated with optical pulses through LD2 with a peak power of 3\,$\mu$W.

\begin{figure}
  \includegraphics[width=\linewidth]{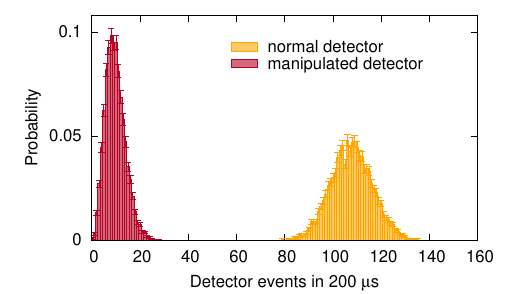}
  \caption{\label{fig:selfseed}
    Distribution of detector events in the presence of self-seeding light in a
    test interval of $T=200\,\mu$s for a normally operating, and
    a manipulated detector. The manipulated detector shows a similar
    distribution as the one in Fig.~\ref{fig:demosetup}(b), while the normally
    operating detector  shows a distinctly higher event number. Error bars
    indicate Poissonian standard deviations resulting from 7432 and 7686 test
    runs for a normal detector and a manipulated detector, respectively. 
    }
 \end{figure}

To demonstrate the first example of detector self-testing, we turn on the
light emitter LE in the test interval $T$ both for a normally operating and a
manipulated detector. The resulting detection event distributions are shown in
Fig.~\ref{fig:selfseed}. For a normally operating detector, the observed APD2
events in the test interval increase significantly to a mean of about
$\bar{n}_{T1}\approx100$, while for a manipulated detector, the distribution is
similar to the ``normal'' distribution with $\bar{n}_{N}\approx10$ in
Fig.~\ref{fig:demosetup}(b). With a threshold at $n=50$, the two distributions
can be easily distinguished, and a detector manipulation attempt
(specifically: the presence of a blinding light level) easily identified
in a single measurement interval $T$; in the experiment, the unmanipulated
detector never showed less than 78 events, while the manipulated showed never
more than 30 events.

\begin{figure}
   \includegraphics[width=\linewidth]{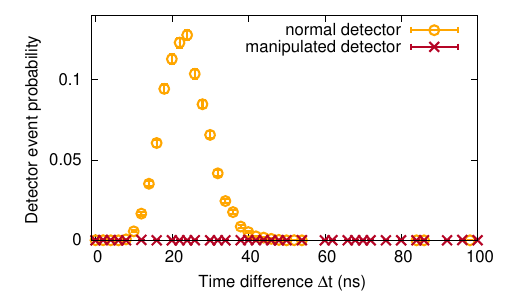}
  \caption{\label{fig:selfseed2}Detector event probability for a 25\,ns long
    bright pulse of the self-testing light emitter LE for a manipulated
    and normal detector vs the time difference $\Delta t$
    between detector event and a self testing pulse edge. A non-manipulated
    detector reacts with an event with high probability within less than
    60\,ns. Optical and 
    electrical delays shift the detector response away from $\Delta t=0$, and
    error bars indicate Poissonian standard deviations resulting from 12542 and 12380 test
    runs for normal detector and manipulated detector, respectively.
  }
\end{figure}

The necessary time to detect a manipulated detector can be shortened even
further with the second example of self-testing. We demonstrate this by
driving the light emitter LE to emit $\delta t=25$\,ns long pulses, and increasing the
coupling to 
the detector APD2 compared to the previous example. Figure~\ref{fig:selfseed2}
shows the probability of registering a signal from APD2 as a function of the
time $\Delta t$ after the start of the self-testing pulse. A non-manipulated
detector shows an overall detector response probability $p_1=93.4\%$ within  60\,ns (11720 photon detection events out of 12542 optical pulse). This
number does not reach 100\%, as the detector may have been in a recovery state
from a previous detection event. For a manipulated detector, i.e., in presence
of both detector blinding and fake states, we find an integral detector event
probability $p_2=0.3\%$ (36 out of 12380 test pulses). These
events were caused by fake states, not by light from the LE. Detector
manipulation (specifically, the detector blinding) can therefore be identified
with a few short test pulses to a very high statistical significance.

\begin{figure}
  \includegraphics[width=\linewidth]{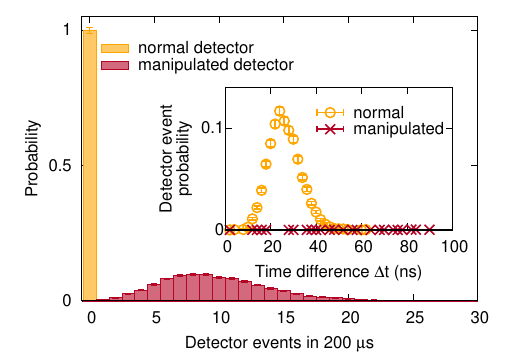}
  \caption{\label{fig:selfblind}Detector event distribution in a test interval
    $T=200\,\mu\text{s}$ in the presence of self-blinding light for a normal
    and manipulated detector, registered 60\,ns {\it after} the onset of
    the self-blinding light. A manipulated detector still reports events due
    to fake states. Inset: probability of a detector
    event in the first 60\,ns after switching on the self-blinding light. This
    scheme allows to detect the presence of both a blinding and fake state
    detector manipulation.
  }
\end{figure}

To demonstrate the third example of detector self-testing, we increased the
optical power of LE on detector APD2 to a level that it could reliably blind
the detector. 
Figure~\ref{fig:selfblind} shows both
a distribution of detection events in a test interval $T=200\,\mu\text{s}$,
taken 60\,ns after the onset of light emission by LE. The
un-manipulated detector is insensitive to single photons in this interval;
we observed only 8 events in 7608 test runs (likely due to electrical noise), while a manipulated detector still reported events due to fake
states present at the input; we observed 7655 out of 7658 events (with the
missing events compatible with statistics).
The onset of the test light emission triggered a
detector reaction within the first 60\,ns with a
probability $p_1=97.6\%$ (7426 detector events out of 7608 test runs, see
inset of Fig.~\ref{fig:selfblind}) for a non-manipulated detector, while the
probability of an onset event was $p_2=0.2\%$ (17 out of 7658 runs) for a
manipulated detector caused by fake states. A local
light emitter that is able to self-blind the detector is thus able to reveal
the presence of both blinding and the fake state in a detector
manipulation attempt.

\textit{Conclusion --}
We demonstrated self-testing of single photon detectors that can reliably
reveal manipulation attacks. The self-testing strategy relies on a light
source near the detector under possible external manipulation, and is
able to detect both negative manipulations (i.e. suppression of
single photon detections) and positive manipulations (i.e., generating
detector events that are not caused by single photon detections) in a
relatively short time with a high statistical significance.
The detector self-testing  makes no assumptions on the nature of the
manipulation attack of the detector, and thus also covers manipulations
that are not of the known nature like detector blinding and fake states.
As the self-testing can be
accomplished by a relatively simple light source (as long as this is outside
the control and knowledge of an adversary), this scheme can address one of the
most significant hardware vulnerabilities of QKD systems in a significantly
simpler way as compared to device-independent or measurement-device independent
approaches, and may even be a suitable to retrofit existing QKD systems
to make them resilient against detector manipulation attacks.

\begin{acknowledgments}
This work was supported by by the National Research Foundation (NRF) Singapore
(Grant: QEP-P1), and through the Research Centres of Excellence programme
supported by NRF Singapore and the Ministry of Education, Singapore.
\end{acknowledgments}


\begin{thebibliography}{37}%
\makeatletter
\providecommand \@ifxundefined [1]{%
 \@ifx{#1\undefined}
}%
\providecommand \@ifnum [1]{%
 \ifnum #1\expandafter \@firstoftwo
 \else \expandafter \@secondoftwo
 \fi
}%
\providecommand \@ifx [1]{%
 \ifx #1\expandafter \@firstoftwo
 \else \expandafter \@secondoftwo
 \fi
}%
\providecommand \natexlab [1]{#1}%
\providecommand \enquote  [1]{``#1''}%
\providecommand \bibnamefont  [1]{#1}%
\providecommand \bibfnamefont [1]{#1}%
\providecommand \citenamefont [1]{#1}%
\providecommand \href@noop [0]{\@secondoftwo}%
\providecommand \href [0]{\begingroup \@sanitize@url \@href}%
\providecommand \@href[1]{\@@startlink{#1}\@@href}%
\providecommand \@@href[1]{\endgroup#1\@@endlink}%
\providecommand \@sanitize@url [0]{\catcode `\\12\catcode `\$12\catcode
  `\&12\catcode `\#12\catcode `\^12\catcode `\_12\catcode `\%12\relax}%
\providecommand \@@startlink[1]{}%
\providecommand \@@endlink[0]{}%
\providecommand \url  [0]{\begingroup\@sanitize@url \@url }%
\providecommand \@url [1]{\endgroup\@href {#1}{\urlprefix }}%
\providecommand \urlprefix  [0]{URL }%
\providecommand \Eprint [0]{\href }%
\providecommand \doibase [0]{http://dx.doi.org/}%
\providecommand \selectlanguage [0]{\@gobble}%
\providecommand \bibinfo  [0]{\@secondoftwo}%
\providecommand \bibfield  [0]{\@secondoftwo}%
\providecommand \translation [1]{[#1]}%
\providecommand \BibitemOpen [0]{}%
\providecommand \bibitemStop [0]{}%
\providecommand \bibitemNoStop [0]{.\EOS\space}%
\providecommand \EOS [0]{\spacefactor3000\relax}%
\providecommand \BibitemShut  [1]{\csname bibitem#1\endcsname}%
\let\auto@bib@innerbib\@empty
\bibitem [{\citenamefont {Bennett}\ and\ \citenamefont
  {Brassard}(1984)}]{BB84}%
  \BibitemOpen
  \bibfield  {author} {\bibinfo {author} {\bibfnamefont {C.~H.}\ \bibnamefont
  {Bennett}}\ and\ \bibinfo {author} {\bibfnamefont {G.}~\bibnamefont
  {Brassard}},\ }\href@noop {} {\bibfield  {journal} {\bibinfo  {journal}
  {Proceedings of IEEE International Conference on Computers, Systems and
  Signal Processing}\ ,\ \bibinfo {pages} {175}} (\bibinfo {year}
  {1984})}\BibitemShut {NoStop}%
\bibitem [{\citenamefont {Ekert}(1991)}]{ekert:91}%
  \BibitemOpen
  \bibfield  {author} {\bibinfo {author} {\bibfnamefont {A.~K.}\ \bibnamefont
  {Ekert}},\ }\href@noop {} {\bibfield  {journal} {\bibinfo  {journal} {Phys.
  Rev. Lett.}\ }\textbf {\bibinfo {volume} {67}},\ \bibinfo {pages} {661}
  (\bibinfo {year} {1991})}\BibitemShut {NoStop}%
\bibitem [{\citenamefont {Gisin}\ \emph {et~al.}(2002)\citenamefont {Gisin},
  \citenamefont {Ribordy}, \citenamefont {Tittel},\ and\ \citenamefont
  {Zbinden}}]{Gisin}%
  \BibitemOpen
  \bibfield  {author} {\bibinfo {author} {\bibfnamefont {N.}~\bibnamefont
  {Gisin}}, \bibinfo {author} {\bibfnamefont {G.}~\bibnamefont {Ribordy}},
  \bibinfo {author} {\bibfnamefont {W.}~\bibnamefont {Tittel}}, \ and\ \bibinfo
  {author} {\bibfnamefont {H.}~\bibnamefont {Zbinden}},\ }\href {\doibase
  10.1103/RevModPhys.74.145} {\bibfield  {journal} {\bibinfo  {journal} {Rev.
  Mod. Phys.}\ }\textbf {\bibinfo {volume} {74}},\ \bibinfo {pages} {145}
  (\bibinfo {year} {2002})}\BibitemShut {NoStop}%
\bibitem [{\citenamefont {Scarani}\ \emph {et~al.}(2009)\citenamefont
  {Scarani}, \citenamefont {Bechmann-Pasquinucci}, \citenamefont {Cerf},
  \citenamefont {Du{\v{s}}ek}, \citenamefont {L{\"{u}}tkenhaus},\ and\
  \citenamefont {Peev}}]{Scarani2009}%
  \BibitemOpen
  \bibfield  {author} {\bibinfo {author} {\bibfnamefont {V.}~\bibnamefont
  {Scarani}}, \bibinfo {author} {\bibfnamefont {H.}~\bibnamefont
  {Bechmann-Pasquinucci}}, \bibinfo {author} {\bibfnamefont {N.~J.}\
  \bibnamefont {Cerf}}, \bibinfo {author} {\bibfnamefont {M.}~\bibnamefont
  {Du{\v{s}}ek}}, \bibinfo {author} {\bibfnamefont {N.}~\bibnamefont
  {L{\"{u}}tkenhaus}}, \ and\ \bibinfo {author} {\bibfnamefont
  {M.}~\bibnamefont {Peev}},\ }\href {\doibase 10.1103/RevModPhys.81.1301}
  {\bibfield  {journal} {\bibinfo  {journal} {Rev. Mod. Phys.}\ }\textbf
  {\bibinfo {volume} {81}},\ \bibinfo {pages} {1301} (\bibinfo {year}
  {2009})}\BibitemShut {NoStop}%
\bibitem [{\citenamefont {Scarani}\ and\ \citenamefont
  {Kurtsiefer}(2014)}]{Scarani2014}%
  \BibitemOpen
  \bibfield  {author} {\bibinfo {author} {\bibfnamefont {V.}~\bibnamefont
  {Scarani}}\ and\ \bibinfo {author} {\bibfnamefont {C.}~\bibnamefont
  {Kurtsiefer}},\ }\href {\doibase 10.1016/j.tcs.2014.09.015} {\bibfield
  {journal} {\bibinfo  {journal} {Theor. Comput. Sci.}\ }\textbf {\bibinfo
  {volume} {560}},\ \bibinfo {pages} {27} (\bibinfo {year} {2014})}\BibitemShut
  {NoStop}%
\bibitem [{\citenamefont {Xu}\ \emph {et~al.}(2020)\citenamefont {Xu},
  \citenamefont {Ma}, \citenamefont {Zhang}, \citenamefont {Lo},\ and\
  \citenamefont {Pan}}]{Xu2020}%
  \BibitemOpen
  \bibfield  {author} {\bibinfo {author} {\bibfnamefont {F.}~\bibnamefont
  {Xu}}, \bibinfo {author} {\bibfnamefont {X.}~\bibnamefont {Ma}}, \bibinfo
  {author} {\bibfnamefont {Q.}~\bibnamefont {Zhang}}, \bibinfo {author}
  {\bibfnamefont {H.~K.}\ \bibnamefont {Lo}}, \ and\ \bibinfo {author}
  {\bibfnamefont {J.~W.}\ \bibnamefont {Pan}},\ }\href {\doibase
  10.1103/REVMODPHYS.92.025002} {\bibfield  {journal} {\bibinfo  {journal}
  {Rev. Mod. Phys.}\ }\textbf {\bibinfo {volume} {92}},\ \bibinfo {pages}
  {025002} (\bibinfo {year} {2020})}\BibitemShut {NoStop}%
\bibitem [{\citenamefont {Brassard}\ \emph {et~al.}(2000)\citenamefont
  {Brassard}, \citenamefont {L{\"{u}}tkenhaus}, \citenamefont {Mor},\ and\
  \citenamefont {Sanders}}]{Brassard2000}%
  \BibitemOpen
  \bibfield  {author} {\bibinfo {author} {\bibfnamefont {G.}~\bibnamefont
  {Brassard}}, \bibinfo {author} {\bibfnamefont {N.}~\bibnamefont
  {L{\"{u}}tkenhaus}}, \bibinfo {author} {\bibfnamefont {T.}~\bibnamefont
  {Mor}}, \ and\ \bibinfo {author} {\bibfnamefont {B.~C.}\ \bibnamefont
  {Sanders}},\ }\href {\doibase 10.1103/PhysRevLett.85.1330} {\bibfield
  {journal} {\bibinfo  {journal} {Phys. Rev. Lett.}\ }\textbf {\bibinfo
  {volume} {85}},\ \bibinfo {pages} {1330} (\bibinfo {year}
  {2000})}\BibitemShut {NoStop}%
\bibitem [{\citenamefont {L{\"{u}}tkenhaus}(2000)}]{Lutkenhaus2000}%
  \BibitemOpen
  \bibfield  {author} {\bibinfo {author} {\bibfnamefont {N.}~\bibnamefont
  {L{\"{u}}tkenhaus}},\ }\href {\doibase 10.1103/PhysRevA.61.052304} {\bibfield
   {journal} {\bibinfo  {journal} {Phys. Rev. A}\ }\textbf {\bibinfo {volume}
  {61}},\ \bibinfo {pages} {10} (\bibinfo {year} {2000})}\BibitemShut {NoStop}%
\bibitem [{\citenamefont {Vakhitov}\ \emph {et~al.}(2001)\citenamefont
  {Vakhitov}, \citenamefont {Makarov},\ and\ \citenamefont
  {Hjelme}}]{Vakhitov2001}%
  \BibitemOpen
  \bibfield  {author} {\bibinfo {author} {\bibfnamefont {A.}~\bibnamefont
  {Vakhitov}}, \bibinfo {author} {\bibfnamefont {V.}~\bibnamefont {Makarov}}, \
  and\ \bibinfo {author} {\bibfnamefont {D.~R.}\ \bibnamefont {Hjelme}},\
  }\href {\doibase 10.1080/09500340108240904} {\bibfield  {journal} {\bibinfo
  {journal} {J. Mod. Opt.}\ }\textbf {\bibinfo {volume} {48}},\ \bibinfo
  {pages} {2023} (\bibinfo {year} {2001})}\BibitemShut {NoStop}%
\bibitem [{\citenamefont {Qi}\ \emph {et~al.}(2007)\citenamefont {Qi},
  \citenamefont {Fung}, \citenamefont {Lo},\ and\ \citenamefont {Ma}}]{Qi2007}%
  \BibitemOpen
  \bibfield  {author} {\bibinfo {author} {\bibfnamefont {B.}~\bibnamefont
  {Qi}}, \bibinfo {author} {\bibfnamefont {C.~H.~F.}\ \bibnamefont {Fung}},
  \bibinfo {author} {\bibfnamefont {H.~K.}\ \bibnamefont {Lo}}, \ and\ \bibinfo
  {author} {\bibfnamefont {X.}~\bibnamefont {Ma}},\ }\href {\doibase
  10.26421/qic7.1-2-3} {\bibfield  {journal} {\bibinfo  {journal} {Quantum Inf.
  Comput.}\ }\textbf {\bibinfo {volume} {7}},\ \bibinfo {pages} {73} (\bibinfo
  {year} {2007})}\BibitemShut {NoStop}%
\bibitem [{\citenamefont {Lamas-Linares}\ and\ \citenamefont
  {Kurtsiefer}(2007)}]{Lamas-Linares2007}%
  \BibitemOpen
  \bibfield  {author} {\bibinfo {author} {\bibfnamefont {A.}~\bibnamefont
  {Lamas-Linares}}\ and\ \bibinfo {author} {\bibfnamefont {C.}~\bibnamefont
  {Kurtsiefer}},\ }\href {\doibase 10.1364/oe.15.009388} {\bibfield  {journal}
  {\bibinfo  {journal} {Opt. Express}\ }\textbf {\bibinfo {volume} {15}},\
  \bibinfo {pages} {9388} (\bibinfo {year} {2007})}\BibitemShut {NoStop}%
\bibitem [{\citenamefont {Zhao}\ \emph {et~al.}(2008)\citenamefont {Zhao},
  \citenamefont {Fung}, \citenamefont {Qi}, \citenamefont {Chen},\ and\
  \citenamefont {Lo}}]{Zhao2008}%
  \BibitemOpen
  \bibfield  {author} {\bibinfo {author} {\bibfnamefont {Y.}~\bibnamefont
  {Zhao}}, \bibinfo {author} {\bibfnamefont {C.~H.~F.}\ \bibnamefont {Fung}},
  \bibinfo {author} {\bibfnamefont {B.}~\bibnamefont {Qi}}, \bibinfo {author}
  {\bibfnamefont {C.}~\bibnamefont {Chen}}, \ and\ \bibinfo {author}
  {\bibfnamefont {H.~K.}\ \bibnamefont {Lo}},\ }\href {\doibase
  10.1103/PhysRevA.78.042333} {\bibfield  {journal} {\bibinfo  {journal} {Phys.
  Rev. A}\ }\textbf {\bibinfo {volume} {78}},\ \bibinfo {pages} {042333}
  (\bibinfo {year} {2008})}\BibitemShut {NoStop}%
\bibitem [{\citenamefont {Makarov}(2009)}]{Makarov2009}%
  \BibitemOpen
  \bibfield  {author} {\bibinfo {author} {\bibfnamefont {V.}~\bibnamefont
  {Makarov}},\ }\href {\doibase 10.1088/1367-2630/11/6/065003} {\bibfield
  {journal} {\bibinfo  {journal} {New J. Phys.}\ }\textbf {\bibinfo {volume}
  {11}},\ \bibinfo {pages} {065003} (\bibinfo {year} {2009})}\BibitemShut
  {NoStop}%
\bibitem [{\citenamefont {Lydersen}\ \emph {et~al.}(2010)\citenamefont
  {Lydersen}, \citenamefont {Wiechers}, \citenamefont {Wittmann}, \citenamefont
  {Elser}, \citenamefont {Skaar},\ and\ \citenamefont
  {Makarov}}]{Lydersen2010}%
  \BibitemOpen
  \bibfield  {author} {\bibinfo {author} {\bibfnamefont {L.}~\bibnamefont
  {Lydersen}}, \bibinfo {author} {\bibfnamefont {C.}~\bibnamefont {Wiechers}},
  \bibinfo {author} {\bibfnamefont {C.}~\bibnamefont {Wittmann}}, \bibinfo
  {author} {\bibfnamefont {D.}~\bibnamefont {Elser}}, \bibinfo {author}
  {\bibfnamefont {J.}~\bibnamefont {Skaar}}, \ and\ \bibinfo {author}
  {\bibfnamefont {V.}~\bibnamefont {Makarov}},\ }\href
  {https://www.nature.com/articles/nphoton.2010.214} {\bibfield  {journal}
  {\bibinfo  {journal} {Nature Photonics}\ }\textbf {\bibinfo {volume} {4}},\
  \bibinfo {pages} {686} (\bibinfo {year} {2010})}\BibitemShut {NoStop}%
\bibitem [{\citenamefont {Lydersen}\ \emph {et~al.}(2011)\citenamefont
  {Lydersen}, \citenamefont {Akhlaghi}, \citenamefont {Majedi}, \citenamefont
  {Skaar},\ and\ \citenamefont {Makarov}}]{Lydersen2011}%
  \BibitemOpen
  \bibfield  {author} {\bibinfo {author} {\bibfnamefont {L.}~\bibnamefont
  {Lydersen}}, \bibinfo {author} {\bibfnamefont {M.~K.}\ \bibnamefont
  {Akhlaghi}}, \bibinfo {author} {\bibfnamefont {A.~H.}\ \bibnamefont
  {Majedi}}, \bibinfo {author} {\bibfnamefont {J.}~\bibnamefont {Skaar}}, \
  and\ \bibinfo {author} {\bibfnamefont {V.}~\bibnamefont {Makarov}},\ }\href
  {\doibase 10.1088/1367-2630/13/11/113042} {\bibfield  {journal} {\bibinfo
  {journal} {New J. Phys.}\ }\textbf {\bibinfo {volume} {13}},\ \bibinfo
  {pages} {113042} (\bibinfo {year} {2011})}\BibitemShut {NoStop}%
\bibitem [{\citenamefont {Goltsman}\ \emph {et~al.}(2019)\citenamefont
  {Goltsman}, \citenamefont {Elezov}, \citenamefont {Ozhegov},\ and\
  \citenamefont {Makarov}}]{Goltsman2019}%
  \BibitemOpen
  \bibfield  {author} {\bibinfo {author} {\bibfnamefont {G.}~\bibnamefont
  {Goltsman}}, \bibinfo {author} {\bibfnamefont {M.}~\bibnamefont {Elezov}},
  \bibinfo {author} {\bibfnamefont {R.}~\bibnamefont {Ozhegov}}, \ and\
  \bibinfo {author} {\bibfnamefont {V.}~\bibnamefont {Makarov}},\ }\href
  {\doibase 10.1364/OE.27.030979} {\bibfield  {journal} {\bibinfo  {journal}
  {Opt. Express, Vol. 27, Issue 21, pp. 30979-30988}\ }\textbf {\bibinfo
  {volume} {27}},\ \bibinfo {pages} {30979} (\bibinfo {year}
  {2019})}\BibitemShut {NoStop}%
\bibitem [{\citenamefont {Gerhardt}\ \emph {et~al.}(2011)\citenamefont
  {Gerhardt}, \citenamefont {Liu}, \citenamefont {Lamas-Linares}, \citenamefont
  {Skaar}, \citenamefont {Kurtsiefer},\ and\ \citenamefont
  {Makarov}}]{Gerhardt2011}%
  \BibitemOpen
  \bibfield  {author} {\bibinfo {author} {\bibfnamefont {I.}~\bibnamefont
  {Gerhardt}}, \bibinfo {author} {\bibfnamefont {Q.}~\bibnamefont {Liu}},
  \bibinfo {author} {\bibfnamefont {A.~A.}\ \bibnamefont {Lamas-Linares}},
  \bibinfo {author} {\bibfnamefont {J.}~\bibnamefont {Skaar}}, \bibinfo
  {author} {\bibfnamefont {C.}~\bibnamefont {Kurtsiefer}}, \ and\ \bibinfo
  {author} {\bibfnamefont {V.}~\bibnamefont {Makarov}},\ }\href {\doibase
  10.1038/ncomms1348} {\bibfield  {journal} {\bibinfo  {journal} {Nat.
  Commun.}\ }\textbf {\bibinfo {volume} {2}},\ \bibinfo {pages} {1} (\bibinfo
  {year} {2011})}\BibitemShut {NoStop}%
\bibitem [{\citenamefont {Honjo}\ \emph {et~al.}(2013)\citenamefont {Honjo},
  \citenamefont {Fujiwara}, \citenamefont {Shimizu}, \citenamefont {Tamaki},
  \citenamefont {Miki}, \citenamefont {Yamashita}, \citenamefont {Terai},
  \citenamefont {Wang},\ and\ \citenamefont {Sasaki}}]{Honjo2013}%
  \BibitemOpen
  \bibfield  {author} {\bibinfo {author} {\bibfnamefont {T.}~\bibnamefont
  {Honjo}}, \bibinfo {author} {\bibfnamefont {M.}~\bibnamefont {Fujiwara}},
  \bibinfo {author} {\bibfnamefont {K.}~\bibnamefont {Shimizu}}, \bibinfo
  {author} {\bibfnamefont {K.}~\bibnamefont {Tamaki}}, \bibinfo {author}
  {\bibfnamefont {S.}~\bibnamefont {Miki}}, \bibinfo {author} {\bibfnamefont
  {T.}~\bibnamefont {Yamashita}}, \bibinfo {author} {\bibfnamefont
  {H.}~\bibnamefont {Terai}}, \bibinfo {author} {\bibfnamefont
  {Z.}~\bibnamefont {Wang}}, \ and\ \bibinfo {author} {\bibfnamefont
  {M.}~\bibnamefont {Sasaki}},\ }\href {\doibase 10.1364/oe.21.002667}
  {\bibfield  {journal} {\bibinfo  {journal} {Opt. Express}\ }\textbf {\bibinfo
  {volume} {21}},\ \bibinfo {pages} {2667} (\bibinfo {year}
  {2013})}\BibitemShut {NoStop}%
\bibitem [{\citenamefont {{Ferreira Da Silva}}\ \emph
  {et~al.}(2015)\citenamefont {{Ferreira Da Silva}}, \citenamefont {{Do
  Amaral}}, \citenamefont {Xavier}, \citenamefont {Temporao},\ and\
  \citenamefont {{Von Der Weid}}}]{FerreiraDaSilva2015}%
  \BibitemOpen
  \bibfield  {author} {\bibinfo {author} {\bibfnamefont {T.}~\bibnamefont
  {{Ferreira Da Silva}}}, \bibinfo {author} {\bibfnamefont {G.~C.}\
  \bibnamefont {{Do Amaral}}}, \bibinfo {author} {\bibfnamefont {G.~B.}\
  \bibnamefont {Xavier}}, \bibinfo {author} {\bibfnamefont {G.~P.}\
  \bibnamefont {Temporao}}, \ and\ \bibinfo {author} {\bibfnamefont {J.~P.}\
  \bibnamefont {{Von Der Weid}}},\ }\href {\doibase 10.1109/JSTQE.2014.2361793}
  {\bibfield  {journal} {\bibinfo  {journal} {IEEE J. Sel. Top. Quantum
  Electron.}\ }\textbf {\bibinfo {volume} {21}},\ \bibinfo {pages} {159}
  (\bibinfo {year} {2015})}\BibitemShut {NoStop}%
\bibitem [{\citenamefont {Wang}\ \emph {et~al.}(2016)\citenamefont {Wang},
  \citenamefont {Wang}, \citenamefont {Qin}, \citenamefont {Wei},\ and\
  \citenamefont {Zhang}}]{Wang2016}%
  \BibitemOpen
  \bibfield  {author} {\bibinfo {author} {\bibfnamefont {J.}~\bibnamefont
  {Wang}}, \bibinfo {author} {\bibfnamefont {H.}~\bibnamefont {Wang}}, \bibinfo
  {author} {\bibfnamefont {X.}~\bibnamefont {Qin}}, \bibinfo {author}
  {\bibfnamefont {Z.}~\bibnamefont {Wei}}, \ and\ \bibinfo {author}
  {\bibfnamefont {Z.}~\bibnamefont {Zhang}},\ }\href {\doibase
  10.1140/epjd/e2015-60469-8} {\bibfield  {journal} {\bibinfo  {journal} {Eur.
  Phys. J. D}\ }\textbf {\bibinfo {volume} {70}},\ \bibinfo {pages} {5}
  (\bibinfo {year} {2016})}\BibitemShut {NoStop}%
\bibitem [{\citenamefont {Gras}\ \emph {et~al.}(2021)\citenamefont {Gras},
  \citenamefont {Rusca}, \citenamefont {Zbinden},\ and\ \citenamefont
  {Bussi{\`{e}}res}}]{Gras2021}%
  \BibitemOpen
  \bibfield  {author} {\bibinfo {author} {\bibfnamefont {G.}~\bibnamefont
  {Gras}}, \bibinfo {author} {\bibfnamefont {D.}~\bibnamefont {Rusca}},
  \bibinfo {author} {\bibfnamefont {H.}~\bibnamefont {Zbinden}}, \ and\
  \bibinfo {author} {\bibfnamefont {F.}~\bibnamefont {Bussi{\`{e}}res}},\
  }\href {\doibase 10.1103/PhysRevApplied.15.034052} {\bibfield  {journal}
  {\bibinfo  {journal} {Phys. Rev. Appl.}\ }\textbf {\bibinfo {volume} {15}},\
  \bibinfo {pages} {034052} (\bibinfo {year} {2021})}\BibitemShut {NoStop}%
\bibitem [{\citenamefont {Dixon}\ \emph {et~al.}(2017)\citenamefont {Dixon},
  \citenamefont {Dynes}, \citenamefont {Lucamarini}, \citenamefont
  {Fr{\"{o}}hlich}, \citenamefont {Sharpe}, \citenamefont {Plews},
  \citenamefont {Tam}, \citenamefont {Yuan}, \citenamefont {Tanizawa},
  \citenamefont {Sato}, \citenamefont {Kawamura}, \citenamefont {Fujiwara},
  \citenamefont {Sasaki},\ and\ \citenamefont {Shields}}]{Dixon2017}%
  \BibitemOpen
  \bibfield  {author} {\bibinfo {author} {\bibfnamefont {A.~R.}\ \bibnamefont
  {Dixon}}, \bibinfo {author} {\bibfnamefont {J.~F.}\ \bibnamefont {Dynes}},
  \bibinfo {author} {\bibfnamefont {M.}~\bibnamefont {Lucamarini}}, \bibinfo
  {author} {\bibfnamefont {B.}~\bibnamefont {Fr{\"{o}}hlich}}, \bibinfo
  {author} {\bibfnamefont {A.~W.}\ \bibnamefont {Sharpe}}, \bibinfo {author}
  {\bibfnamefont {A.}~\bibnamefont {Plews}}, \bibinfo {author} {\bibfnamefont
  {W.}~\bibnamefont {Tam}}, \bibinfo {author} {\bibfnamefont {Z.~L.}\
  \bibnamefont {Yuan}}, \bibinfo {author} {\bibfnamefont {Y.}~\bibnamefont
  {Tanizawa}}, \bibinfo {author} {\bibfnamefont {H.}~\bibnamefont {Sato}},
  \bibinfo {author} {\bibfnamefont {S.}~\bibnamefont {Kawamura}}, \bibinfo
  {author} {\bibfnamefont {M.}~\bibnamefont {Fujiwara}}, \bibinfo {author}
  {\bibfnamefont {M.}~\bibnamefont {Sasaki}}, \ and\ \bibinfo {author}
  {\bibfnamefont {A.~J.}\ \bibnamefont {Shields}},\ }\href {\doibase
  10.1038/s41598-017-01884-0} {\bibfield  {journal} {\bibinfo  {journal} {Sci.
  Rep.}\ }\textbf {\bibinfo {volume} {7}},\ \bibinfo {pages} {1} (\bibinfo
  {year} {2017})}\BibitemShut {NoStop}%
\bibitem [{\citenamefont {Lim}\ \emph {et~al.}(2015)\citenamefont {Lim},
  \citenamefont {Walenta}, \citenamefont {Legr{\'{e}}}, \citenamefont {Gisin},\
  and\ \citenamefont {Zbinden}}]{Lim2015}%
  \BibitemOpen
  \bibfield  {author} {\bibinfo {author} {\bibfnamefont {C.~C.~W.}\
  \bibnamefont {Lim}}, \bibinfo {author} {\bibfnamefont {N.}~\bibnamefont
  {Walenta}}, \bibinfo {author} {\bibfnamefont {M.}~\bibnamefont
  {Legr{\'{e}}}}, \bibinfo {author} {\bibfnamefont {N.}~\bibnamefont {Gisin}},
  \ and\ \bibinfo {author} {\bibfnamefont {H.}~\bibnamefont {Zbinden}},\ }\href
  {\doibase 10.1109/JSTQE.2015.2389528} {\bibfield  {journal} {\bibinfo
  {journal} {IEEE J. Sel. Top. Quantum Electron.}\ }\textbf {\bibinfo {volume}
  {21}},\ \bibinfo {pages} {192} (\bibinfo {year} {2015})}\BibitemShut
  {NoStop}%
\bibitem [{\citenamefont {Qian}\ \emph {et~al.}(2019)\citenamefont {Qian},
  \citenamefont {He}, \citenamefont {Wang}, \citenamefont {Chen}, \citenamefont
  {Yin}, \citenamefont {Guo},\ and\ \citenamefont {Han}}]{Qian2019}%
  \BibitemOpen
  \bibfield  {author} {\bibinfo {author} {\bibfnamefont {Y.-J.}\ \bibnamefont
  {Qian}}, \bibinfo {author} {\bibfnamefont {D.-Y.}\ \bibnamefont {He}},
  \bibinfo {author} {\bibfnamefont {S.}~\bibnamefont {Wang}}, \bibinfo {author}
  {\bibfnamefont {W.}~\bibnamefont {Chen}}, \bibinfo {author} {\bibfnamefont
  {Z.-Q.}\ \bibnamefont {Yin}}, \bibinfo {author} {\bibfnamefont {G.-C.}\
  \bibnamefont {Guo}}, \ and\ \bibinfo {author} {\bibfnamefont {Z.-F.}\
  \bibnamefont {Han}},\ }\href {\doibase 10.1364/optica.6.001178} {\bibfield
  {journal} {\bibinfo  {journal} {Optica}\ }\textbf {\bibinfo {volume} {6}},\
  \bibinfo {pages} {1178} (\bibinfo {year} {2019})}\BibitemShut {NoStop}%
\bibitem [{\citenamefont {Yuan}\ \emph {et~al.}(2010)\citenamefont {Yuan},
  \citenamefont {Dynes},\ and\ \citenamefont {Shields}}]{Yuan2010}%
  \BibitemOpen
  \bibfield  {author} {\bibinfo {author} {\bibfnamefont {Z.}~\bibnamefont
  {Yuan}}, \bibinfo {author} {\bibfnamefont {J.~F.}\ \bibnamefont {Dynes}}, \
  and\ \bibinfo {author} {\bibfnamefont {A.~J.}\ \bibnamefont {Shields}},\
  }\href@noop {} {\bibfield  {journal} {\bibinfo  {journal} {Nature Photonics}\
  }\textbf {\bibinfo {volume} {4}},\ \bibinfo {pages} {800} (\bibinfo {year}
  {2010})}\BibitemShut {NoStop}%
\bibitem [{\citenamefont {Yuan}\ \emph {et~al.}(2011)\citenamefont {Yuan},
  \citenamefont {Dynes},\ and\ \citenamefont {Shields}}]{Yuan2011}%
  \BibitemOpen
  \bibfield  {author} {\bibinfo {author} {\bibfnamefont {Z.~L.}\ \bibnamefont
  {Yuan}}, \bibinfo {author} {\bibfnamefont {J.~F.}\ \bibnamefont {Dynes}}, \
  and\ \bibinfo {author} {\bibfnamefont {A.~J.}\ \bibnamefont {Shields}},\
  }\href {\doibase 10.1063/1.3597221} {\bibfield  {journal} {\bibinfo
  {journal} {Appl. Phys. Lett.}\ }\textbf {\bibinfo {volume} {98}},\ \bibinfo
  {pages} {231104} (\bibinfo {year} {2011})}\BibitemShut {NoStop}%
\bibitem [{\citenamefont {Huang}\ \emph {et~al.}(2016)\citenamefont {Huang},
  \citenamefont {Sajeed}, \citenamefont {Chaiwongkhot}, \citenamefont
  {Soucarros}, \citenamefont {Legr{\'{e}}},\ and\ \citenamefont
  {Makarov}}]{Cnt-cnt-Makarov}%
  \BibitemOpen
  \bibfield  {author} {\bibinfo {author} {\bibfnamefont {A.}~\bibnamefont
  {Huang}}, \bibinfo {author} {\bibfnamefont {S.}~\bibnamefont {Sajeed}},
  \bibinfo {author} {\bibfnamefont {P.}~\bibnamefont {Chaiwongkhot}}, \bibinfo
  {author} {\bibfnamefont {M.}~\bibnamefont {Soucarros}}, \bibinfo {author}
  {\bibfnamefont {M.}~\bibnamefont {Legr{\'{e}}}}, \ and\ \bibinfo {author}
  {\bibfnamefont {V.}~\bibnamefont {Makarov}},\ }\href {\doibase
  10.1109/JQE.2016.2611443} {\bibfield  {journal} {\bibinfo  {journal} {IEEE
  Journal of Quantum Electronics}\ }\textbf {\bibinfo {volume} {52}} (\bibinfo
  {year} {2016}),\ 10.1109/JQE.2016.2611443}\BibitemShut {NoStop}%
\bibitem [{\citenamefont {Lo}\ \emph {et~al.}(2012)\citenamefont {Lo},
  \citenamefont {Curty},\ and\ \citenamefont {Qi}}]{MDIQKD:2012}%
  \BibitemOpen
  \bibfield  {author} {\bibinfo {author} {\bibfnamefont {H.-K.}\ \bibnamefont
  {Lo}}, \bibinfo {author} {\bibfnamefont {M.}~\bibnamefont {Curty}}, \ and\
  \bibinfo {author} {\bibfnamefont {B.}~\bibnamefont {Qi}},\ }\href {\doibase
  10.1103/PhysRevLett.108.130503} {\bibfield  {journal} {\bibinfo  {journal}
  {Phys. Rev. Lett.}\ }\textbf {\bibinfo {volume} {108}},\ \bibinfo {pages}
  {130503} (\bibinfo {year} {2012})}\BibitemShut {NoStop}%
\bibitem [{\citenamefont {Pironio}\ \emph {et~al.}(2009)\citenamefont
  {Pironio}, \citenamefont {Ac{\'{\i}}n}, \citenamefont {Brunner},
  \citenamefont {Gisin}, \citenamefont {Massar},\ and\ \citenamefont
  {Scarani}}]{DIQKD:2009}%
  \BibitemOpen
  \bibfield  {author} {\bibinfo {author} {\bibfnamefont {S.}~\bibnamefont
  {Pironio}}, \bibinfo {author} {\bibfnamefont {A.}~\bibnamefont
  {Ac{\'{\i}}n}}, \bibinfo {author} {\bibfnamefont {N.}~\bibnamefont
  {Brunner}}, \bibinfo {author} {\bibfnamefont {N.}~\bibnamefont {Gisin}},
  \bibinfo {author} {\bibfnamefont {S.}~\bibnamefont {Massar}}, \ and\ \bibinfo
  {author} {\bibfnamefont {V.}~\bibnamefont {Scarani}},\ }\href {\doibase
  10.1088/1367-2630/11/4/045021} {\bibfield  {journal} {\bibinfo  {journal}
  {New Journal of Physics}\ }\textbf {\bibinfo {volume} {11}},\ \bibinfo
  {pages} {045021} (\bibinfo {year} {2009})}\BibitemShut {NoStop}%
\bibitem [{\citenamefont {Liu}\ \emph {et~al.}(2013)\citenamefont {Liu},
  \citenamefont {Chen}, \citenamefont {Wang}, \citenamefont {Liang},
  \citenamefont {Shentu}, \citenamefont {Wang}, \citenamefont {Cui},
  \citenamefont {Yin}, \citenamefont {Liu}, \citenamefont {Li}, \citenamefont
  {Ma}, \citenamefont {Pelc}, \citenamefont {Fejer}, \citenamefont {Peng},
  \citenamefont {Zhang},\ and\ \citenamefont {Pan}}]{MDIexp1}%
  \BibitemOpen
  \bibfield  {author} {\bibinfo {author} {\bibfnamefont {Y.}~\bibnamefont
  {Liu}}, \bibinfo {author} {\bibfnamefont {T.-Y.}\ \bibnamefont {Chen}},
  \bibinfo {author} {\bibfnamefont {L.-J.}\ \bibnamefont {Wang}}, \bibinfo
  {author} {\bibfnamefont {H.}~\bibnamefont {Liang}}, \bibinfo {author}
  {\bibfnamefont {G.-L.}\ \bibnamefont {Shentu}}, \bibinfo {author}
  {\bibfnamefont {J.}~\bibnamefont {Wang}}, \bibinfo {author} {\bibfnamefont
  {K.}~\bibnamefont {Cui}}, \bibinfo {author} {\bibfnamefont {H.-L.}\
  \bibnamefont {Yin}}, \bibinfo {author} {\bibfnamefont {N.-L.}\ \bibnamefont
  {Liu}}, \bibinfo {author} {\bibfnamefont {L.}~\bibnamefont {Li}}, \bibinfo
  {author} {\bibfnamefont {X.}~\bibnamefont {Ma}}, \bibinfo {author}
  {\bibfnamefont {J.~S.}\ \bibnamefont {Pelc}}, \bibinfo {author}
  {\bibfnamefont {M.~M.}\ \bibnamefont {Fejer}}, \bibinfo {author}
  {\bibfnamefont {C.-Z.}\ \bibnamefont {Peng}}, \bibinfo {author}
  {\bibfnamefont {Q.}~\bibnamefont {Zhang}}, \ and\ \bibinfo {author}
  {\bibfnamefont {J.-W.}\ \bibnamefont {Pan}},\ }\href {\doibase
  10.1103/PhysRevLett.111.130502} {\bibfield  {journal} {\bibinfo  {journal}
  {Physical Review Letters}\ }\textbf {\bibinfo {volume} {111}},\ \bibinfo
  {pages} {130502} (\bibinfo {year} {2013})}\BibitemShut {NoStop}%
\bibitem [{\citenamefont {Tang}\ \emph {et~al.}(2014)\citenamefont {Tang},
  \citenamefont {Yin}, \citenamefont {Chen}, \citenamefont {Liu}, \citenamefont
  {Zhang}, \citenamefont {Jiang}, \citenamefont {Zhang}, \citenamefont {Wang},
  \citenamefont {You}, \citenamefont {Guan}, \citenamefont {Yang},
  \citenamefont {Wang}, \citenamefont {Liang}, \citenamefont {Zhang},
  \citenamefont {Zhou}, \citenamefont {Ma}, \citenamefont {Chen}, \citenamefont
  {Zhang},\ and\ \citenamefont {Pan}}]{MDI200km}%
  \BibitemOpen
  \bibfield  {author} {\bibinfo {author} {\bibfnamefont {Y.-L.}\ \bibnamefont
  {Tang}}, \bibinfo {author} {\bibfnamefont {H.-L.}\ \bibnamefont {Yin}},
  \bibinfo {author} {\bibfnamefont {S.-J.}\ \bibnamefont {Chen}}, \bibinfo
  {author} {\bibfnamefont {Y.}~\bibnamefont {Liu}}, \bibinfo {author}
  {\bibfnamefont {W.-J.}\ \bibnamefont {Zhang}}, \bibinfo {author}
  {\bibfnamefont {X.}~\bibnamefont {Jiang}}, \bibinfo {author} {\bibfnamefont
  {L.}~\bibnamefont {Zhang}}, \bibinfo {author} {\bibfnamefont
  {J.}~\bibnamefont {Wang}}, \bibinfo {author} {\bibfnamefont {L.-X.}\
  \bibnamefont {You}}, \bibinfo {author} {\bibfnamefont {J.-Y.}\ \bibnamefont
  {Guan}}, \bibinfo {author} {\bibfnamefont {D.-X.}\ \bibnamefont {Yang}},
  \bibinfo {author} {\bibfnamefont {Z.}~\bibnamefont {Wang}}, \bibinfo {author}
  {\bibfnamefont {H.}~\bibnamefont {Liang}}, \bibinfo {author} {\bibfnamefont
  {Z.}~\bibnamefont {Zhang}}, \bibinfo {author} {\bibfnamefont
  {N.}~\bibnamefont {Zhou}}, \bibinfo {author} {\bibfnamefont {X.}~\bibnamefont
  {Ma}}, \bibinfo {author} {\bibfnamefont {T.-Y.}\ \bibnamefont {Chen}},
  \bibinfo {author} {\bibfnamefont {Q.}~\bibnamefont {Zhang}}, \ and\ \bibinfo
  {author} {\bibfnamefont {J.-W.}\ \bibnamefont {Pan}},\ }\href {\doibase
  10.1103/PhysRevLett.113.190501} {\bibfield  {journal} {\bibinfo  {journal}
  {Physical Review Letters}\ }\textbf {\bibinfo {volume} {113}},\ \bibinfo
  {pages} {190501} (\bibinfo {year} {2014})}\BibitemShut {NoStop}%
\bibitem [{\citenamefont {Yin}\ \emph {et~al.}(2016)\citenamefont {Yin},
  \citenamefont {Chen}, \citenamefont {Yu}, \citenamefont {Liu}, \citenamefont
  {You}, \citenamefont {Zhou}, \citenamefont {Chen}, \citenamefont {Mao},
  \citenamefont {Huang}, \citenamefont {Zhang}, \citenamefont {Chen},
  \citenamefont {Li}, \citenamefont {Nolan}, \citenamefont {Zhou},
  \citenamefont {Jiang}, \citenamefont {Wang}, \citenamefont {Zhang},
  \citenamefont {Wang},\ and\ \citenamefont {Pan}}]{MDI400km}%
  \BibitemOpen
  \bibfield  {author} {\bibinfo {author} {\bibfnamefont {H.-L.}\ \bibnamefont
  {Yin}}, \bibinfo {author} {\bibfnamefont {T.-Y.}\ \bibnamefont {Chen}},
  \bibinfo {author} {\bibfnamefont {Z.-W.}\ \bibnamefont {Yu}}, \bibinfo
  {author} {\bibfnamefont {H.}~\bibnamefont {Liu}}, \bibinfo {author}
  {\bibfnamefont {L.-X.}\ \bibnamefont {You}}, \bibinfo {author} {\bibfnamefont
  {Y.-H.}\ \bibnamefont {Zhou}}, \bibinfo {author} {\bibfnamefont {S.-J.}\
  \bibnamefont {Chen}}, \bibinfo {author} {\bibfnamefont {Y.}~\bibnamefont
  {Mao}}, \bibinfo {author} {\bibfnamefont {M.-Q.}\ \bibnamefont {Huang}},
  \bibinfo {author} {\bibfnamefont {W.-J.}\ \bibnamefont {Zhang}}, \bibinfo
  {author} {\bibfnamefont {H.}~\bibnamefont {Chen}}, \bibinfo {author}
  {\bibfnamefont {M.~J.}\ \bibnamefont {Li}}, \bibinfo {author} {\bibfnamefont
  {D.}~\bibnamefont {Nolan}}, \bibinfo {author} {\bibfnamefont
  {F.}~\bibnamefont {Zhou}}, \bibinfo {author} {\bibfnamefont {X.}~\bibnamefont
  {Jiang}}, \bibinfo {author} {\bibfnamefont {Z.}~\bibnamefont {Wang}},
  \bibinfo {author} {\bibfnamefont {Q.}~\bibnamefont {Zhang}}, \bibinfo
  {author} {\bibfnamefont {X.-B.}\ \bibnamefont {Wang}}, \ and\ \bibinfo
  {author} {\bibfnamefont {J.-W.}\ \bibnamefont {Pan}},\ }\href {\doibase
  10.1103/PhysRevLett.117.190501} {\bibfield  {journal} {\bibinfo  {journal}
  {Physical Review Letters}\ }\textbf {\bibinfo {volume} {117}},\ \bibinfo
  {pages} {190501} (\bibinfo {year} {2016})}\BibitemShut {NoStop}%
\bibitem [{\citenamefont {Liu}\ \emph {et~al.}(2019)\citenamefont {Liu},
  \citenamefont {Wang}, \citenamefont {Wei}, \citenamefont {Fang},
  \citenamefont {Li}, \citenamefont {Liu}, \citenamefont {Liang}, \citenamefont
  {Zhang}, \citenamefont {Zhang}, \citenamefont {Li}, \citenamefont {You},
  \citenamefont {Wang}, \citenamefont {Lo}, \citenamefont {Chen}, \citenamefont
  {Xu},\ and\ \citenamefont {Pan}}]{MDIasym}%
  \BibitemOpen
  \bibfield  {author} {\bibinfo {author} {\bibfnamefont {H.}~\bibnamefont
  {Liu}}, \bibinfo {author} {\bibfnamefont {W.}~\bibnamefont {Wang}}, \bibinfo
  {author} {\bibfnamefont {K.}~\bibnamefont {Wei}}, \bibinfo {author}
  {\bibfnamefont {X.-T.}\ \bibnamefont {Fang}}, \bibinfo {author}
  {\bibfnamefont {L.}~\bibnamefont {Li}}, \bibinfo {author} {\bibfnamefont
  {N.-L.}\ \bibnamefont {Liu}}, \bibinfo {author} {\bibfnamefont
  {H.}~\bibnamefont {Liang}}, \bibinfo {author} {\bibfnamefont {S.-J.}\
  \bibnamefont {Zhang}}, \bibinfo {author} {\bibfnamefont {W.}~\bibnamefont
  {Zhang}}, \bibinfo {author} {\bibfnamefont {H.}~\bibnamefont {Li}}, \bibinfo
  {author} {\bibfnamefont {L.}~\bibnamefont {You}}, \bibinfo {author}
  {\bibfnamefont {Z.}~\bibnamefont {Wang}}, \bibinfo {author} {\bibfnamefont
  {H.-K.}\ \bibnamefont {Lo}}, \bibinfo {author} {\bibfnamefont {T.-Y.}\
  \bibnamefont {Chen}}, \bibinfo {author} {\bibfnamefont {F.}~\bibnamefont
  {Xu}}, \ and\ \bibinfo {author} {\bibfnamefont {J.-W.}\ \bibnamefont {Pan}},\
  }\href {\doibase 10.1103/PhysRevLett.122.160501} {\bibfield  {journal}
  {\bibinfo  {journal} {Physical Review Letters}\ }\textbf {\bibinfo {volume}
  {122}},\ \bibinfo {pages} {160501} (\bibinfo {year} {2019})}\BibitemShut
  {NoStop}%
\bibitem [{\citenamefont {Mayers}\ and\ \citenamefont
  {Yao}(1998)}]{Mayers1998QuantumCW}%
  \BibitemOpen
  \bibfield  {author} {\bibinfo {author} {\bibfnamefont {D.}~\bibnamefont
  {Mayers}}\ and\ \bibinfo {author} {\bibfnamefont {A.~C.-C.}\ \bibnamefont
  {Yao}},\ }\href@noop {} {\bibfield  {journal} {\bibinfo  {journal}
  {Proceedings 39th Annual Symposium on Foundations of Computer Science (Cat.
  No.98CB36280)}\ ,\ \bibinfo {pages} {503}} (\bibinfo {year}
  {1998})}\BibitemShut {NoStop}%
\bibitem [{\citenamefont {van Dam}\ \emph {et~al.}(2000)\citenamefont {van
  Dam}, \citenamefont {Magniez}, \citenamefont {Mosca},\ and\ \citenamefont
  {Santha}}]{vanDam:2000}%
  \BibitemOpen
  \bibfield  {author} {\bibinfo {author} {\bibfnamefont {W.}~\bibnamefont {van
  Dam}}, \bibinfo {author} {\bibfnamefont {F.}~\bibnamefont {Magniez}},
  \bibinfo {author} {\bibfnamefont {M.}~\bibnamefont {Mosca}}, \ and\ \bibinfo
  {author} {\bibfnamefont {M.}~\bibnamefont {Santha}},\ }in\ \href {\doibase
  10.1145/335305.335402} {\emph {\bibinfo {booktitle} {Proceedings of the
  Thirty-Second Annual ACM Symposium on Theory of Computing}}},\ \bibinfo
  {series and number} {STOC '00}\ (\bibinfo  {publisher} {Association for
  Computing Machinery},\ \bibinfo {address} {New York, NY, USA},\ \bibinfo
  {year} {2000})\ p.\ \bibinfo {pages} {688696}\BibitemShut {NoStop}%
\bibitem [{\citenamefont {{\v{S}}upi{\'{c}}}\ and\ \citenamefont
  {Bowles}(2020)}]{Supic2020selftestingof}%
  \BibitemOpen
  \bibfield  {author} {\bibinfo {author} {\bibfnamefont {I.}~\bibnamefont
  {{\v{S}}upi{\'{c}}}}\ and\ \bibinfo {author} {\bibfnamefont {J.}~\bibnamefont
  {Bowles}},\ }\href {\doibase 10.22331/q-2020-09-30-337} {\bibfield  {journal}
  {\bibinfo  {journal} {{Quantum}}\ }\textbf {\bibinfo {volume} {4}},\ \bibinfo
  {pages} {337} (\bibinfo {year} {2020})}\BibitemShut {NoStop}%
\bibitem [{\citenamefont {Tanner}\ \emph {et~al.}(2014)\citenamefont {Tanner},
  \citenamefont {Hadfield},\ and\ \citenamefont {Makarov}}]{Tanner2014}%
  \BibitemOpen
  \bibfield  {author} {\bibinfo {author} {\bibfnamefont {M.~G.}\ \bibnamefont
  {Tanner}}, \bibinfo {author} {\bibfnamefont {R.~H.}\ \bibnamefont
  {Hadfield}}, \ and\ \bibinfo {author} {\bibfnamefont {V.}~\bibnamefont
  {Makarov}},\ }\href {\doibase 10.1364/OE.22.006734} {\bibfield  {journal}
  {\bibinfo  {journal} {Opt. Express, Vol. 22, Issue 6, pp. 6734-6748}\
  }\textbf {\bibinfo {volume} {22}},\ \bibinfo {pages} {6734} (\bibinfo {year}
  {2014})}\BibitemShut {NoStop}%
\end{thebibliography}
\end{document}